\documentclass[final,3p,times,twocolumn]{elsarticle}

\usepackage{amssymb}
\usepackage{color}
\usepackage{subfigure}
\usepackage{url}

\usepackage[switch]{lineno}

\usepackage{diagbox}
\usepackage[colorlinks,linkcolor=blue,anchorcolor=blue,citecolor=blue]{hyperref}
\biboptions{sort&compress}
\usepackage{threeparttable}

\journal{NIM A}

\begin{document}

\begin{frontmatter}



\title{Leakage Current Simulations of Low Gain Avalanche Diode with Improved Radiation Damage Modeling}

\author[label1,label3]{Tao Yang}
\author[label1,label3]{Kewei Wu}
\author[label1,label2]{Mei Zhao}
\author[label1,label3]{Xuewei Jia}
\author[label1,label3]{Yuhang Tan}
\author[label1,label3]{Suyu Xiao}
\author[label1]{Kai Liu}
\author[label1]{Xiyuan Zhang}
\author[label1]{Congcong Wang}

\author[label1,label3]{Mengzhao Li}
\author[label1]{Yunyun Fan}
\author[label1,label3]{Shuqi Li}
\author[label1,label3]{Chengjun Yu}
\author[label1,label3]{Han Cui}

\author[label1,label3]{Hao Zeng}
\author[label1,label3]{Mingjie Zhai}
\author[label1,label3]{Shuiting Xin}
\author[label1,label3]{Maoqiang Jing}

\author[label4]{Gangping Yan}
\author[label4]{Qionghua Zhai}
\author[label4]{Mingzheng Ding}
\author[label4]{Gaobo Xu}
\author[label4]{Huaxiang Yin}

\author[label5]{Gregor Kramberger}

\author[label1,label2]{Zhijun Liang}
\author[label1,label2]{Jo\~{a}o Guimar\~{a}es da Costa}
\author[label1,label2]{Xin Shi\corref{cor1}}
\ead{shixin@ihep.ac.cn}

\address[label1]{Institute of High Energy Physics, Chinese Academy of Sciences, 19B Yuquan Road, Shijingshan District, Beijing 100049, China}
\address[label2]{State Key Laboratory of Particle Detection and Electronics, 19B Yuquan Road, Shijingshan District, Beijing 100049, China}
\address[label3]{University of Chinese Academy of Sciences, 19A Yuquan Road, Shijingshan District, Beijing 100049, China}
\address[label4]{Institute of Microelectronics, Chinese Academy of Sciences, 3rd Beitucheng West Road, Chaoyang District, Beijing 100029, China}
\address[label5]{Jozef Stefan Institute, SI-1000 Ljubljana, Slovenia}
\cortext[cor1]{Corresponding author}

\begin{abstract}
We report precise TCAD simulations of IHEP-IME-v1 Low Gain Avalanche Diode (LGAD) calibrated by secondary ion mass spectroscopy (SIMS). Our setup allows us to evaluate the leakage current, capacitance, and breakdown voltage of LGAD, which agree with measurements' results before irradiation. And we propose an improved LGAD Radiation Damage Model (LRDM) which combines local acceptor removal with global deep energy levels. The LRDM is applied to the IHEP-IME-v1 LGAD and able to predict the leakage current well at -30 $^{\circ}$C  after an irradiation fluence of $ \Phi_{eq}=2.5 \times 10^{15} ~n_{eq}/cm^{2}$. The charge collection efficiency (CCE) is under development.

\end{abstract}



\begin{keyword}
LGAD \sep TCAD simulation \sep acceptor removal \sep neutron irradiation \sep radiation model


\end{keyword}

\end{frontmatter}



\section{Introduction} \label{sec:introduction}
Low Gain Avalanche Diodes (LGADs) are characterized by an excellent timing 
performance with a time resolution of at least 50 ps and have been developed extensively in the past few years \cite{CNM_LGAD,HPK_LGAD,FBK_LGAD,YunyunFAN_NDL,SuyuXIAO_TestBeam,YuhanTAN_NDL_CIAE,BNL_LGAD,Yuzhen_NDL_33um}, especially by the CERN RD50 Collaboration \cite{RD50} in view of the High Luminosity Large Hadron Collider (HL-LHC) upgrade at CERN. 

LGADs will be used in ATLAS and CMS experiments at the LHC. In particular, they will be used place in the High Granularity Timing Detector (HGTD) \cite{ATLAS_HGTD} of ATLAs and in the End-cap Timing Layer (ETL) \cite{CMS_ETL} of CMS. Radiation hardness up to $ 2.5\times 10^{15} ~n_{eq}/cm^{2}$ is required by ATLAS and represents the main challenge in the development of LGAD sensors. Technology Computer-Aided Design (TCAD) is usually employed in the optimization of semiconductor processing and for predicting device performance. In particular, TCAD is useful to accurately predict LGAD sensor features before and after irradiation and, in turn, to understand the mechanisms behind radiation damage in LGADs.
    
A major goal in the design of LGAD sensors is to control the breakdown voltage ($ V_{BD}$) at an appropriate level due to the high electric field near the gain layer that easily exceeds the threshold from linear-model to geiger-model. A longer range of operating voltage with higher $ V_{BD}$ could obtain an appropriate gain, meanwhile ensuring carrier velocity saturation when the device works. In this framework, numerical simulations based on TCAD tools are useful in assessing the performance of calibrated devices before manufacturing, in turn, tuning the sought-after breakdown voltage. 

Previously TCAD analysis of $V_{BD}$ of LGAD sensors \cite{LGAD_VBD_Sim_With_SIMS,CNM_LGAD_VBD_Sim} have shown that $V_{BD}$ strongly depends on the gain layer doping profile and the avalanche model. Calibrated process simulation with SIMS measurement may be used to minimize the influence of doping uncertainty, whereas using appropriate physical models allows one to predict the threshold of breakdown more precisely. In this paper, we report results obtained by TCAD simulations of the full process of IHEP-IME-v1 LGADs. The good agreement between TCAD simulations and experimental measurements 
confirms that this kind of analysis will be a resource to optimize the LGAD performance in the next production stage. It also paves the way for establishing the solid ground for simulations of irradiation processes.

Furthermore, including previous phenomenological studies to build a better model for irradiated LGAD sensors and improve predictions of TCAD simulations is an important topic in itself \cite{RD50}. In previous studies \cite{LGAD_Rad_Sim_Silvaco,LGAD_Rad_Sim_Sentaurus}, different models have been applied to TCAD simulations to obtain qualitative results for the behavior of irradiated LGAD sensors. The results from a comprehensive study \cite{LGAD_Rad_Gain_Sim} analyzing the gain factor has been verified on irradiated LGAD sensors from different vendors. The leakage current and capacitance of irradiated LGAD sensor obtained in \cite{LGAD_Rad_Sim_RT_1,LGAD_Rad_Sim_RT_2,LGAD_Rad_Sim_RT_3} show good agreement with simulated results at room temperature (T=27$ ^{\circ}C$). In this paper, the details and performance of IHEP-IME-v1 LGAD sensors are introduced in Sec.~\ref{sec:ihep_ime_v1}. Then we present the results of precise TCAD simulations based on IHEP-IME-v1 LGAD sensors in Sec.~\ref{sec:tcad_cal}, and introduce a TCAD-LRDM model to evaluate the leakage current and the capacitance of irradiated LGAD sensors at low temperature (T=-30$ ^{\circ}C$) in Sec.~\ref{sec:rad_model}. The leakage current obtained from simulations are in good agreement with experimental measurements, indicating the validity of our model, which combines the acceptor removal with the deep energy levels.

    \section{Devices under study} \label{sec:ihep_ime_v1}

    \begin{figure*}[htb] 
        \centering
        \includegraphics[scale=0.3]{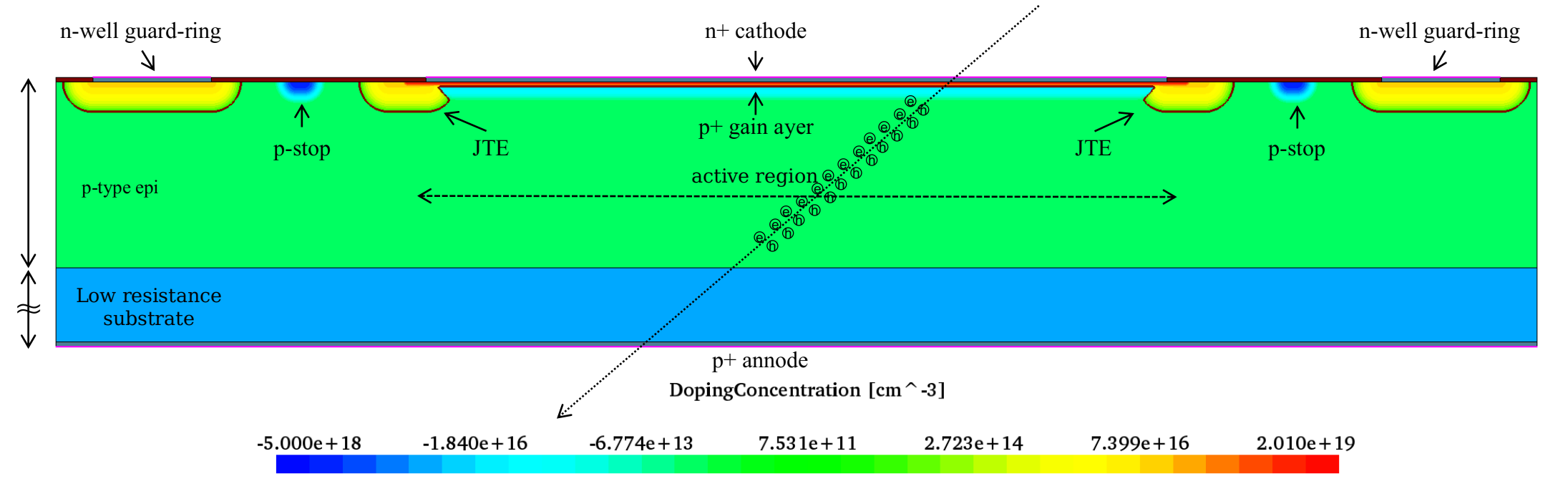}
        \caption{Schematic cross-section and doping concentration of the LGAD studied in this work, 
        with illustration of JTE, P-stop, and N-well guard-ring.}
        \label{fig:lgad_section}
    \end{figure*}

The IHEP-IME-v1 LGAD sensor is being developed by the Institute of High Energy Physics and the Institute of Microelectronics of the Chinese Academy of Sciences \cite{KeweiWu_LGAD}. A cross-section of IHEP-IME-v1 LGAD structure is shown in Fig.\ref{fig:lgad_section}. Typically,the structure contains a (n++)-(p+)-p-(p++) stack, a junction termination extension (JTE), a p-stop and a guard-ring. JTE is included to possibly prevent premature breakdown \cite{HPK_LGAD}. Guard-ring may be regarded as a ``cathode'' for the peripheral region, and it simultaneously reshapes electric field distribution near the region of JTE. P-stop blocks up the potential connection between cathode and guard-ring, due to JTE diffusion process and interface charges after radiation when the same voltage is applied to cathode/guard-ring. It is manufactured on a 50 $ \mu m$ high resistivity p-type epi-wafer with $ \sim$725 $\mu m$ supporting substrate. 
The detailed wafer information is reported in \tablename~\ref{tab:wafer_spec}. The electric field peak in the p+ gain layer is larger than 300 kV/cm when the LGAD sensor operates at nominal bias voltage. In this condition, the drift electrons (a few holes) acquire sufficient energy to generate electron-hole pairs by impact ionization. However, they are still lower than the avalanche threshold. The gain of the LGAD sensor is kept under control at around 10-60, since in this condition good time resolution $ \sigma_{t}$ and signal to noise ratio $ S/N$ may be achieved simultaneously, thanks to the carriers multiplication and the thin active region~\cite{LGAD_Theory}.
    
    \begin{figure}[htb]
        \centering
        \subfigure[]{ \label{fig:l1_15_100}
        \includegraphics[scale=0.25]{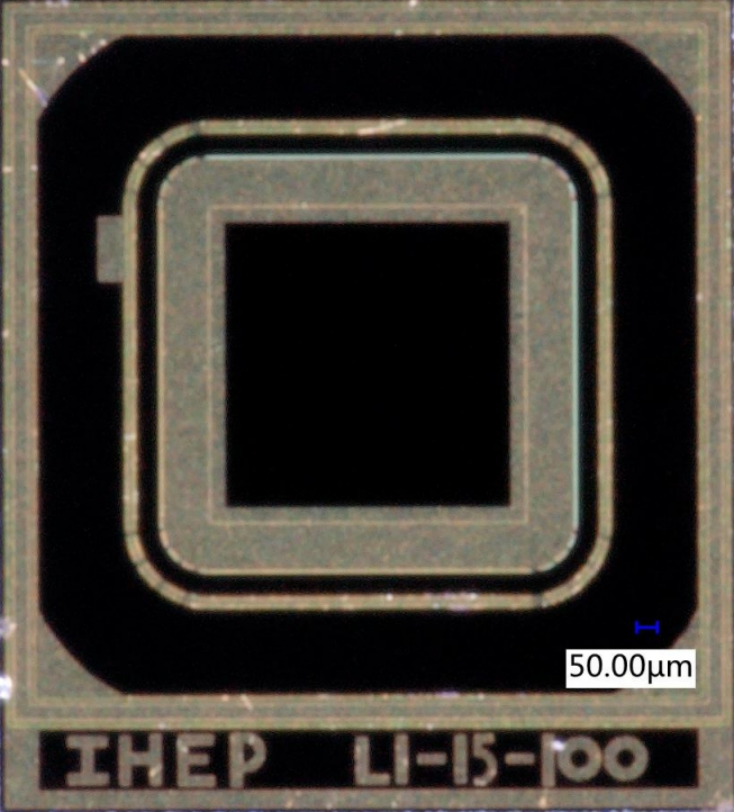}}  
        \subfigure[]{ \label{fig:pin_15_100}
        \includegraphics[scale=0.25]{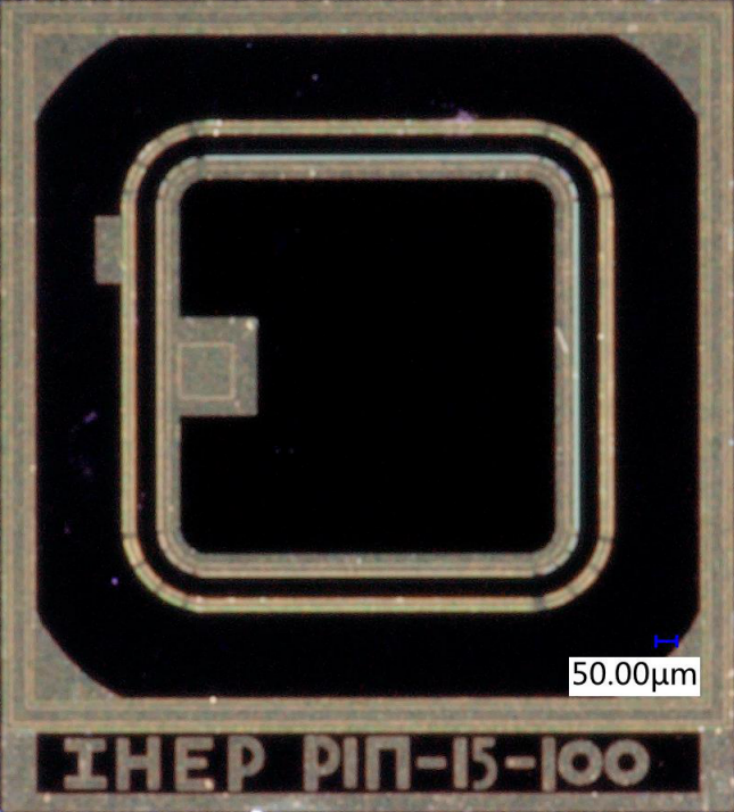}}
        \caption{(a) Top view of IHEP-IME-v1 LGAD; (b) Top view of the PIN devices with same JTE and guard-ring structures. The inner metal-ring is the n+ contact and the outer one is the guard-ring contact.}
    \end{figure}

\begin{figure*}[htb]
        \centering
        \subfigure[]{ \label{fig:lgad_pin_iv}
        \includegraphics[scale=0.33]{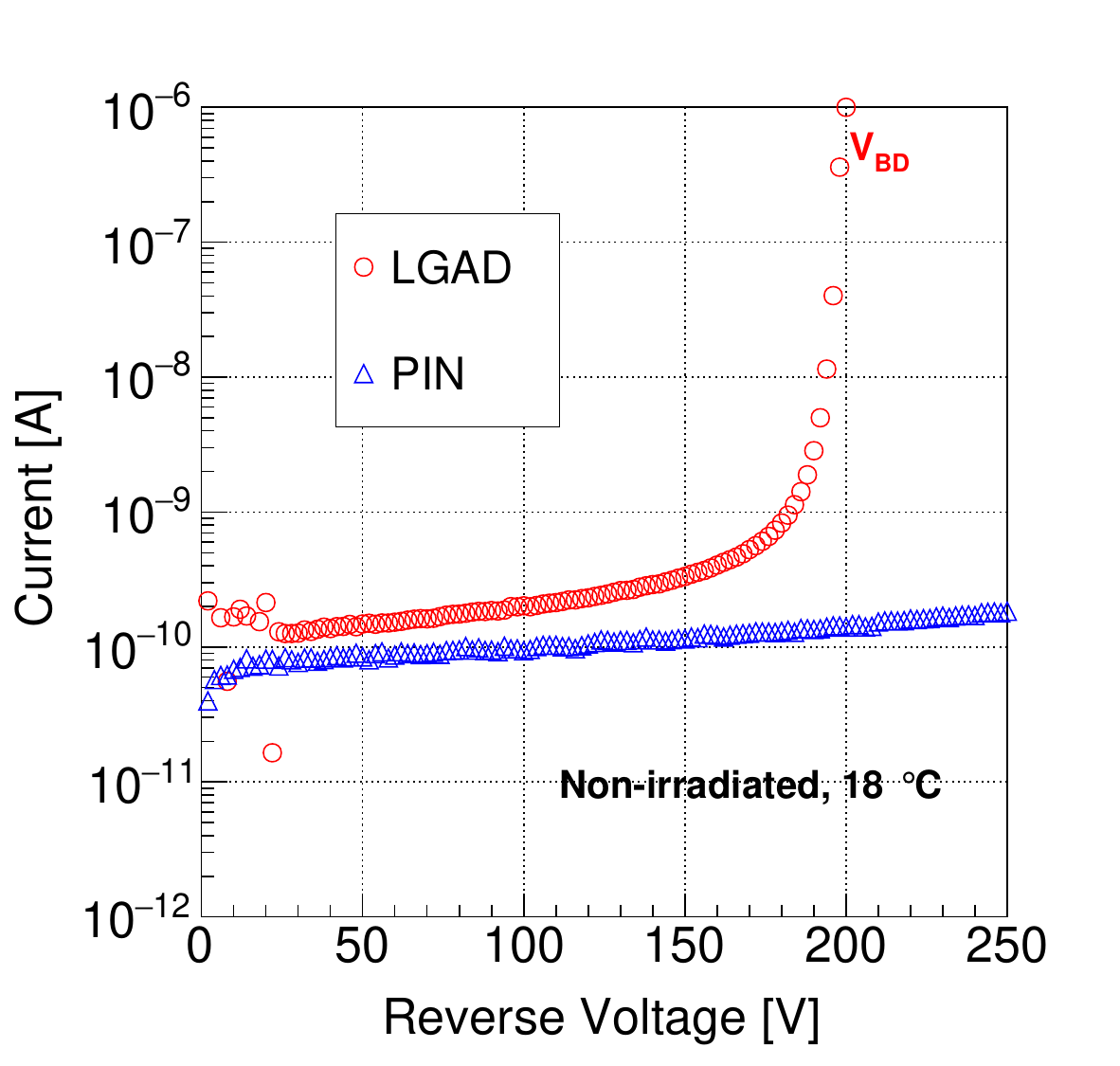}}  
        \subfigure[]{ \label{fig:lgad_pin_cv}
        \includegraphics[scale=0.33]{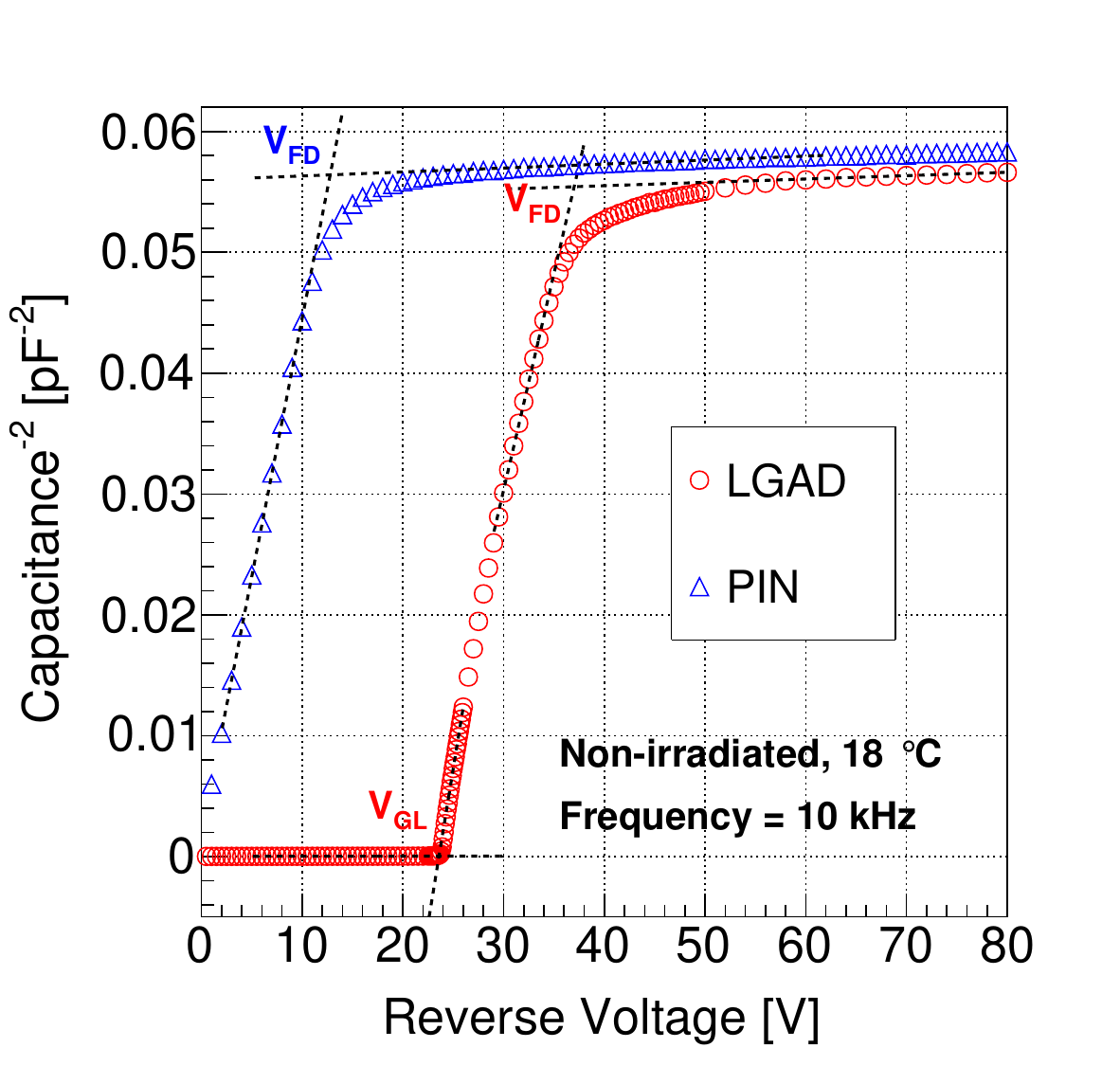}}
        \caption{(a) I-V and (b) C-V characteristic curves of non-irradiated IHEP-IME-v1 LGAD and PIN devices measured at $ 18\pm2$ $^{\circ}$C and $ 45\%$ humidity. The $ V_{GL}$ and $ V_{FD}$ of LGAD are obtained by fitting two linear functions.}
    \end{figure*}

To understand the difference between measurements and simulations, we select two fabricated devices from the IHEP-IME-v1 production line: LGAD (L1-15-100) and PIN (PIN-15-100). They have the same geometry and fabricating process except from the fact that the PIN has no gain layer (see Fig.\ref{fig:l1_15_100} and \ref{fig:pin_15_100}). The active area of both is $ 1.3\times1.3~mm^{2}$. Fig.\ref{fig:lgad_pin_iv} and \ref{fig:lgad_pin_cv} show the I-V and 1/C$^{2}$-V characteristic curves of LGAD/PIN, measured at room temperature. Owing to the presence of an internal gain, the leakage current of the LGAD sensor can be as high as 15 times the leakage current of PIN. The $ V_{BD}$ of LGAD ($\sim$196 V), defined as the voltage corresponding to a leakage current exceeding $ 10^{-6}~A$, is smaller than PIN diode which does not show breakdown up to 700V. The change in capacitance between gain layer depletion voltage ($ V_{GL}$) and full depleted voltage ($ V_{FD}$) in LGAD sensor corresponds to the presence of the (n++)-(p+)-p-(p++) stack.

    \begin{table}[]
        \caption{wafer specification} 
        
        \centering
        \begin{tabular}{p{40pt}p{50pt}p{70pt}}
            \hline
            \hline
                    & parameter     & value                     \\
            \hline
            Epitaxy     &  thickness    & $ 50\sim55 ~\mu m$         \\
                     &  uniformity$^{*}$      & 6\%                       \\
                     &  dopant          & Boron                \\
                     &  resistivity  & $ >1000 ~\Omega \cdot cm$ \\
            \hline
            Substrate     &  material     & CZ silicon                \\
                     &  thickness   & $ 725\pm 20 ~\mu m$            \\
                     &  orientation  & $<$100$>$                    \\
                     &  dopant          & Boron                     \\
                     &  resistivity  & $ < 0.02  ~\Omega \cdot cm$   \\    
            \hline       
        \end{tabular}
        \begin{tablenotes}
            \scriptsize
            \item $^{*}$ Thickness uniformity measured by 9 points.
        \end{tablenotes}
        \label{tab:wafer_spec}
    \end{table}

The effective doping concentration $ N_{eff}$ can be obained from $ d(1/C^{2})/dV$ as \cite{Ar_Model}

    \begin{equation} \label{eq:n_eff}
         N_{eff} = \frac{2}{q \varepsilon A^{2}d(1/C^{2})/dV}
    \end{equation}

\noindent where $ q$ is the electron charge, $ \varepsilon$ is the dielectric constant of silicon and $ A$ is the area of the active region. The effective doping of the bulk region ($ N_{bulk} = 9.8 \times 10^{12}~cm^{-3}$) is obtained from the slope of 1/C$^{2}$-V curve before full depletion of the PIN. The corresponding resistivity $ 1355 ~\Omega \cdot cm$ is consistent with the resistivity of the epitaxial layer provided by the wafer foundry in \tablename~\ref{tab:wafer_spec}. It should be emphasized that equation (\ref{eq:n_eff}) is accurate if the depleted region grows from the main junction side, a situation usually referred to as the   ``one-sided junction" approximation. The same method may be applied to LGAD, but the less accurate determination of doping profile such as shifting of depth is obtained(see Fig.\ref{fig:sims}).

\section{TCAD simulation} \label{sec:tcad_cal}

The full TCAD simulation includes device construction and the simulation of its electrical properties. The device construction involves the definition of geometry and doping profile. The geometry may be scaled-down to reduce the time burden and the doping profile may be extracted from SIMS or generated by process simulation. For the simulation of the electrical properties, the selected physical models should consider the influence of high electric field and doping concentration level in LGAD. More details of physical models are in Sec.~\ref{sec:physical_model}.

\subsection{Doping profile from SIMS}

\begin{figure}[htb]
        \centering 
        \includegraphics[scale=0.33]{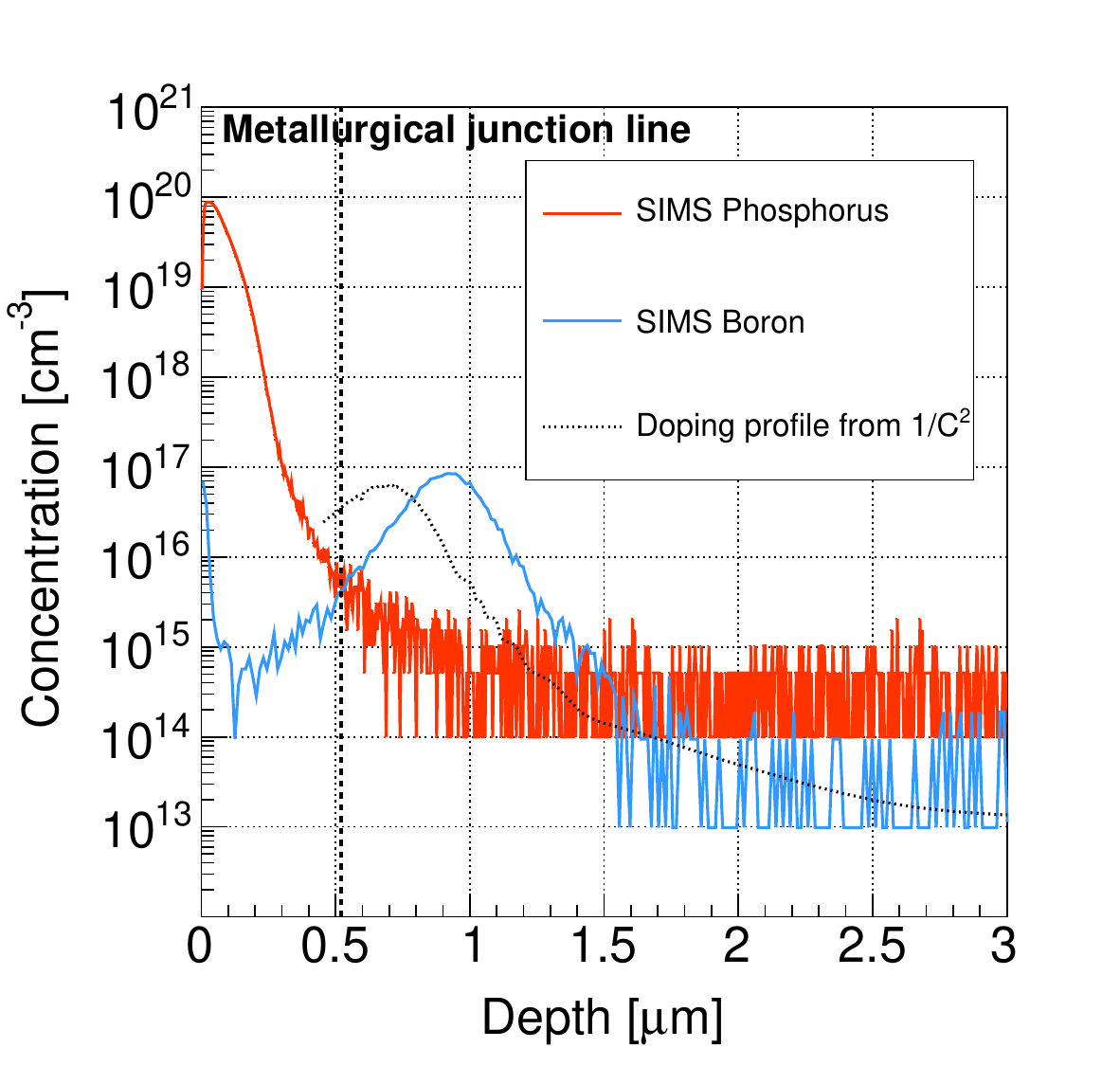}
        \caption{(Color online) SIMS results for IHEP-IME-v1 LGAD sensor: phosphorus(red line), boron(blue line), calculated effective doping from 1/C$^2$-V (dashed line). The detection limit of SIMS technology are: phosphorus $\sim 10^{15} ~atom/cm^{3}$, 
        boron $\sim 10^{14} ~atom/cm^{3}$.}
        \label{fig:sims}
    \end{figure}

Fig.\ref{fig:sims} shows the phosphorus and boron concentrations measured by SIMS. The maximum electric field value appears on the metallurgical junction line. The low energy and extremely high dose of implanted phosphorus lead to a sharp and shallow Ohmic contact made by a n+ layer. The deep p+ gain layer is built by high energy boron implanting where the peak concentration is about $8\times10^{16}~ cm^{-3}$. To avoid damages to the wafer surface during the fabrication process of IHEP-IME-v1 LGADs, which is due to the high energy of ions, a thermal screen oxide layer ($ \sim$$ 120~nm$) covers the wafer surface before boron implantation. The influence of this screen oxide layer on the doping profile is discussed in the next section.

\subsection{Full process calibration by SIMS} \label{subsec:process_cal}   
    
\begin{figure*}[htb]
        \centering
        \subfigure[]{ \label{fig_sims_tcad_p}
        \includegraphics[scale=0.38]{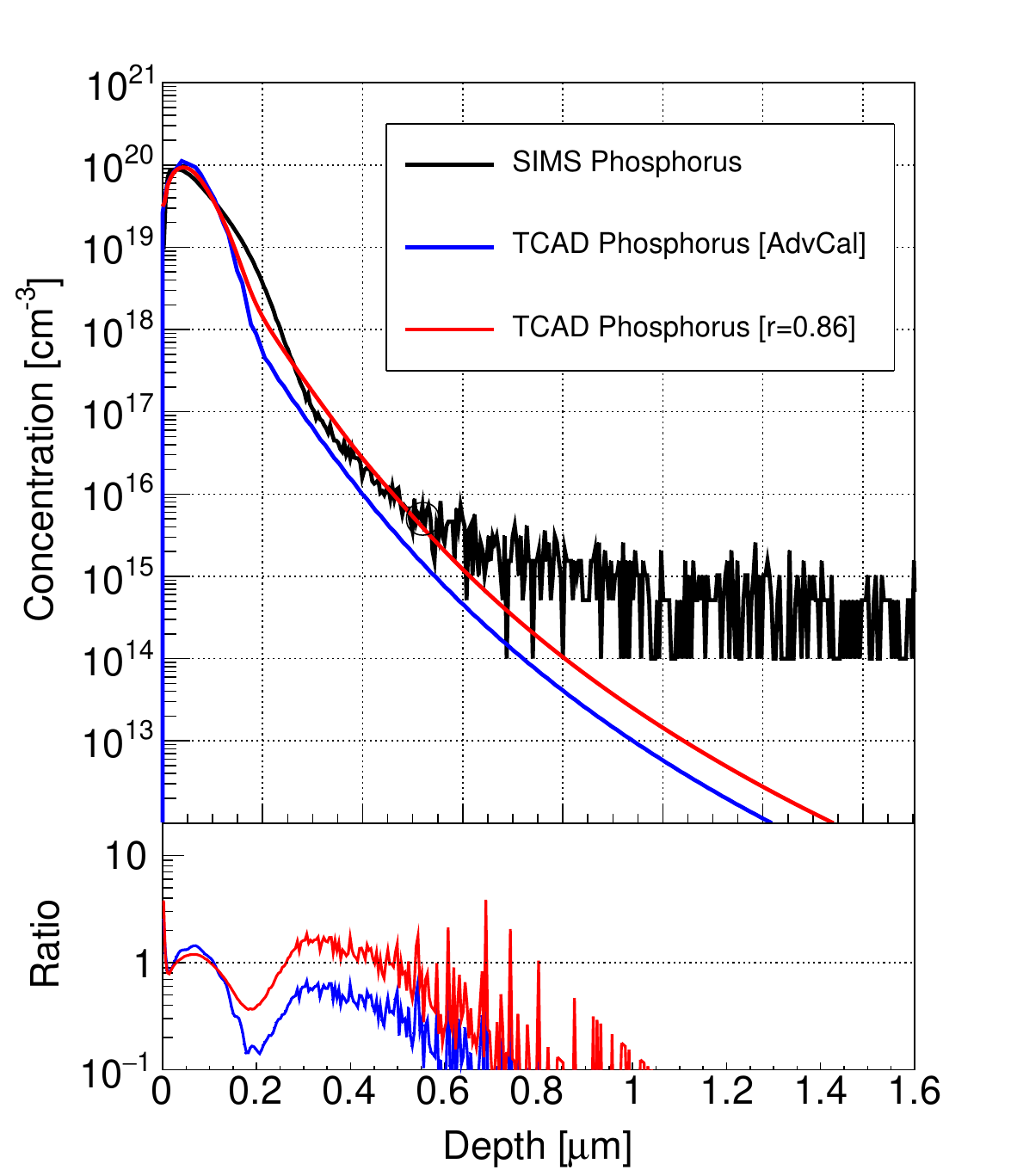}} 
        \subfigure[]{ \label{fig_sims_tcad_b}
        \includegraphics[scale=0.38]{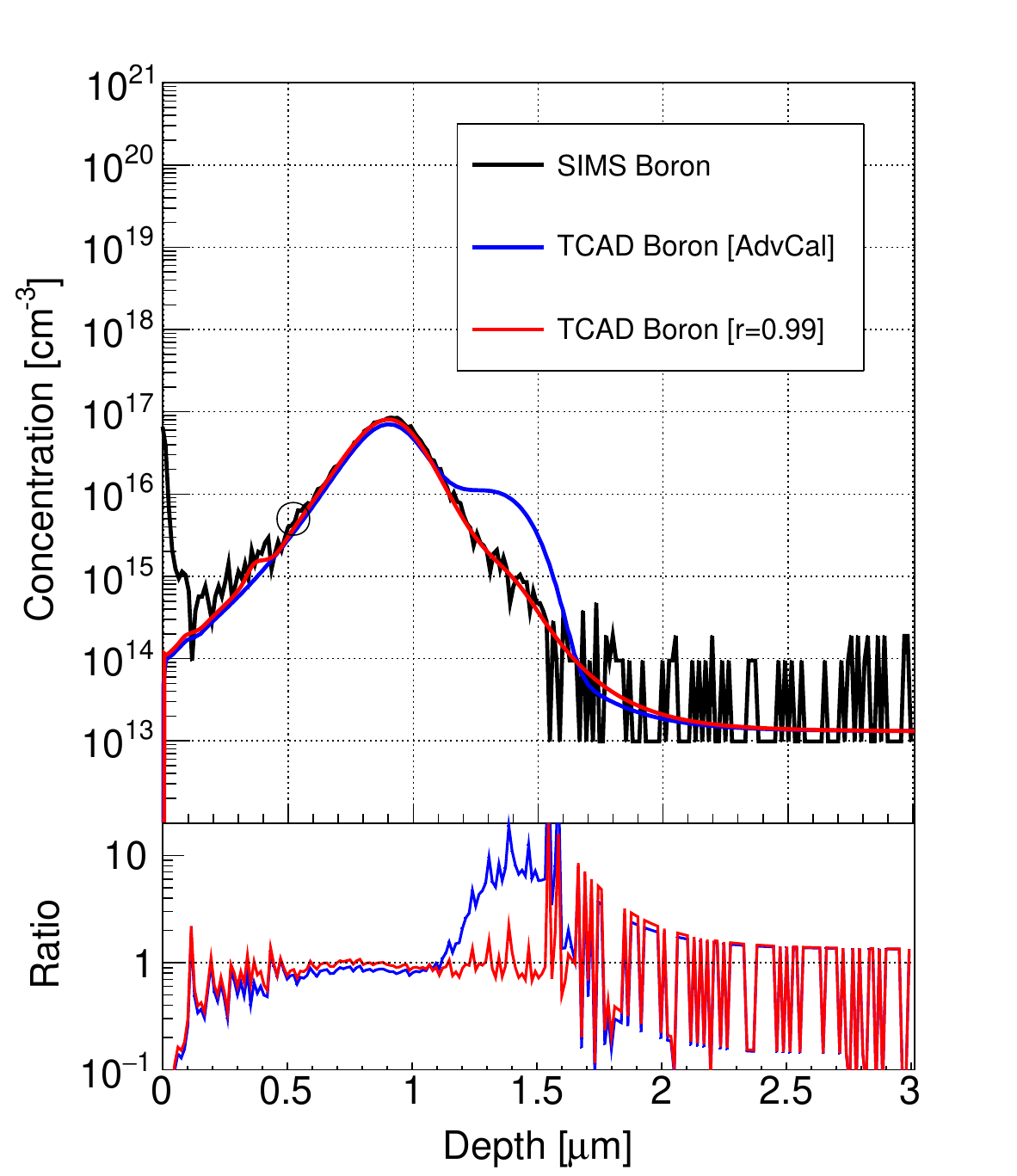}}
    
        \caption{(Color online) (a) phosphorus and (b) boron distribution before and after calibration. The simulated doping profiles are consistent with SIMS results after adjusting the ratio  $r$ between the amorphous and channeling dose. There is a huge distortion between the simulated doping profiles by ``AdvancedCalibration'' and SIMS results. The circle marks the position of the metallurgical junction line. The bottom plots are the corresponding ratio between simulation and SIMS results.}
    \end{figure*}

The simulation process needs to be calibrated by SIMS data before being able to provide a prediction of electrical features. The TCAD environment provides a calibrated database referred to  as ``Advanced Calibration'' \cite{TCAD_Manual}, but there are large discrepancies between simulation based on this database and SIMS data (see Fig.\ref{fig_sims_tcad_p} and \ref{fig_sims_tcad_b}). For phosphorus, the simulated doping is shallower than that obtained from SIMS. For boron, there is an additional bump in the tail region, indicating that the channeling effects are overestimated in the simulations. The possible reason for this behavior may be found in the channeling effects of the screen oxide layer mentioned in the last section. Although we use the full process simulation and implant the boron/phosphorus after growing the screen oxide layer, the effects are not suppressed. In order to calibrate the ion implantation step in our simulations, the ratio $r$ between the amorphous and channeling doses \cite{TCAD_Manual} should be corrected to match our working conditions.

For a data-driven calibration, the ratio of {``channeling effects suppression''} are adjusted for both phosphorus and boron to minimize $ (N_{TCAD} - N_{SIMS})^2$. The optimal values are  $ r_{boron} = 0.99$ and $ r_{phosphorus} = 0.86$ . After calibration, the simulated doping profiles are closer with SIMS data than default.

\subsection{The physical model used in TCAD} \label{sec:physical_model}
The doping dependence, high field saturation on mobility, and temperature dependence on Shockley–Read–Hall Recombination (SRH) are considered, because the effective doping is larger than $ 10^{16} ~cm^{-3}$ , the electric field in the gain layer is about $ 4 \times 10^5 ~V/cm$ and device works in a low-temperature condition. The avalanche is described by a van Overstraeten – de Man model \cite{vOv_Impact_Model} with ``parallel'' driving force and a bandgap dependence. In the van Overstraeten – de Man model, there are two different sets of parameters to describe the carrier generation coefficient that determines the breakdown threshold of LGAD sensor, depending on whether the electric field is in the range from $ 1.75 \times 10^5 ~V/cm$ to $ 4\times 10^5 ~V/cm$ or it is larger.

\subsection{Simulations of the I-V \& C-V characteristics of non-irradiated LGAD sensors}

    \begin{figure*}[htb]
        \centering
        \subfigure[]{ \label{fig:lgad_iv_tcad}
        \includegraphics[scale=0.33]{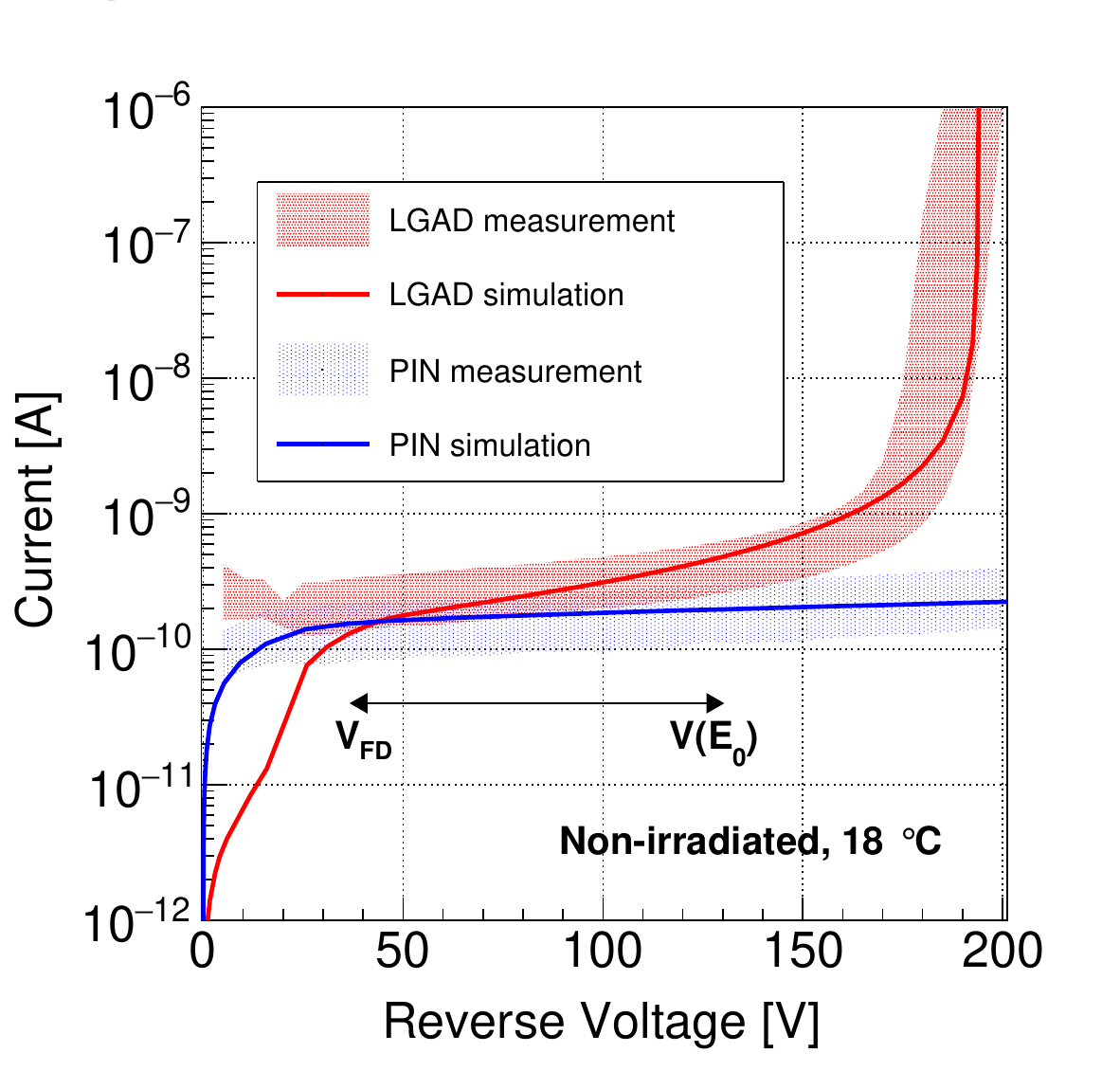}} 
        \subfigure[]{ \label{fig:lgad_cv_tcad}
        \includegraphics[scale=0.33]{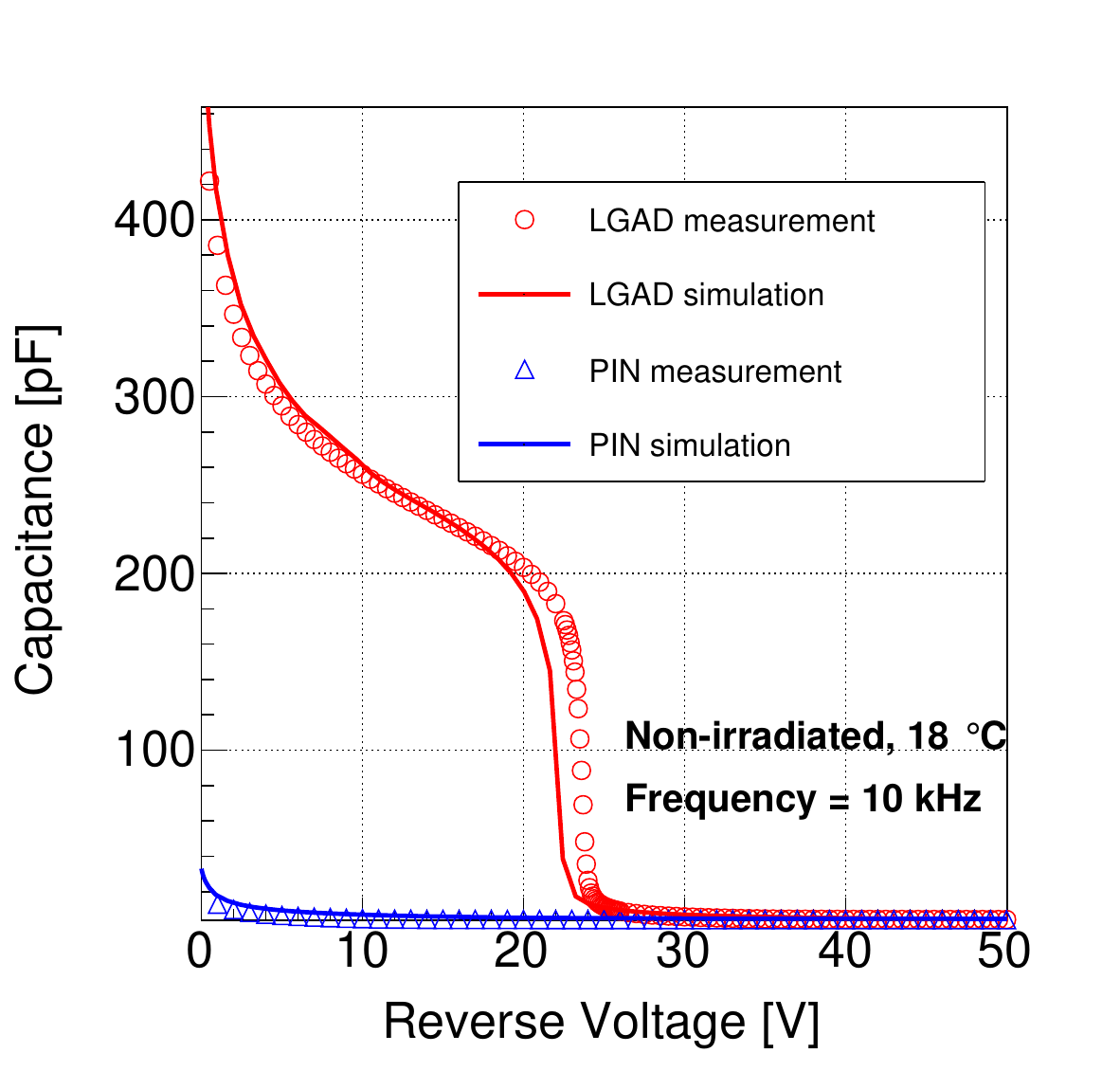}}
        \caption{(Color online) (a) I-V and (b) C-V for un-irradiated LGAD and PIN. The color band of measured I-V is from several LGAD devices which have same structure and fabricated process.
        }
    \end{figure*}

\begin{figure}[htb]
        \centering
        \includegraphics[scale=0.33]{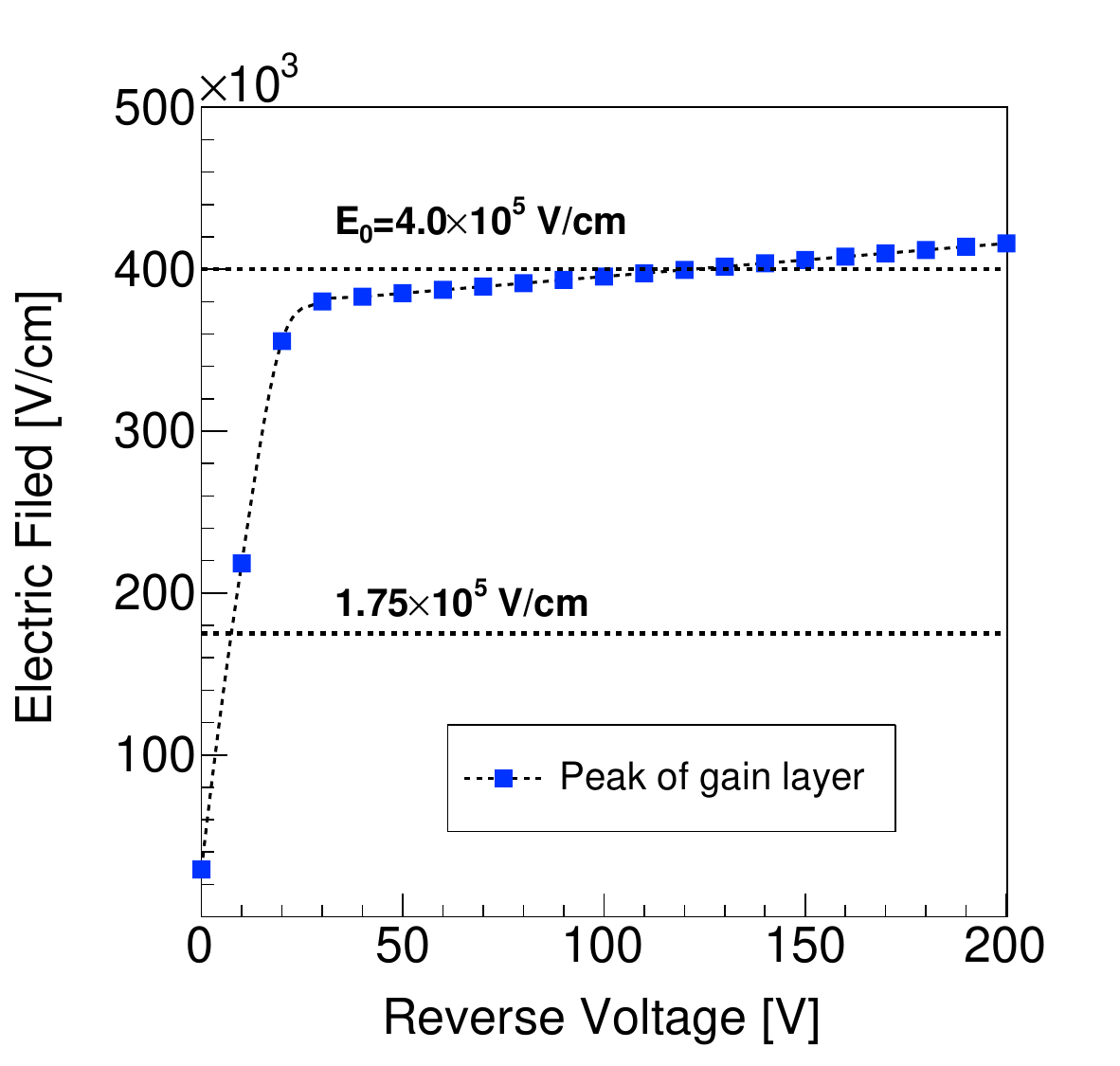}
        \caption{Simulated electric field peak at the metallurgical junction line in un-irradiated LGAD. The  two additional dash lines denote $ 1.75 \times 10^5 V/cm$ and $ 4\times 10^5 V/cm$ electric field threshold for van Overstraeten model. }
        \label{fig:lgad_e_peak}
    \end{figure}

We use PINs and LGADs in order to verify our device simulation setup, and assess the calibrated process simulation as well as the physical model. Several devices are 
subject to measurement to estimate the intrinsic spread of the I-V characteristics due to 
the fabrication process and the measurement conditions (denoted by a colored band in Fig.\ref{fig:lgad_iv_tcad}). The simulated current agrees well with measurements. In particular, simulations reliably predict the breakdown voltage of LGAD sensors. The bias region from $ V_{FD}$ to $ V(E_{0})$, denoted by the arrow in Fig.\ref{fig:lgad_iv_tcad}, is the operation region with moderate gain. $ V(E_{0})$ is the voltage when the maximum electric field at the metallurgical junction line reaches $E_{0}=4\times 10^5 ~V/cm$. The predicted breakdown voltage agrees with the prediction from the van Overstraeten model. In Fig.\ref{fig:lgad_e_peak}, we show the electric field peak at the metallurgical junction line ($\sim0.5~\mu m$, see \figurename~\ref{fig:sims}), which is simulated by TCAD using the calibrated process described in Section~\ref{subsec:process_cal}. When the electric field exceeds the threshold $ E_{0}$, leakage current increases sharply due to the generation of a higher field. The simulated capacitances are shown in Fig.\ref{fig:lgad_cv_tcad}, and are in good agreement with the measured values. The depleted capacitances of simulation and measurement at 50~V both are 5~pF which is compatible with device geometry.

\begin{figure}[htb]
        \centering
        \includegraphics[scale=0.33]{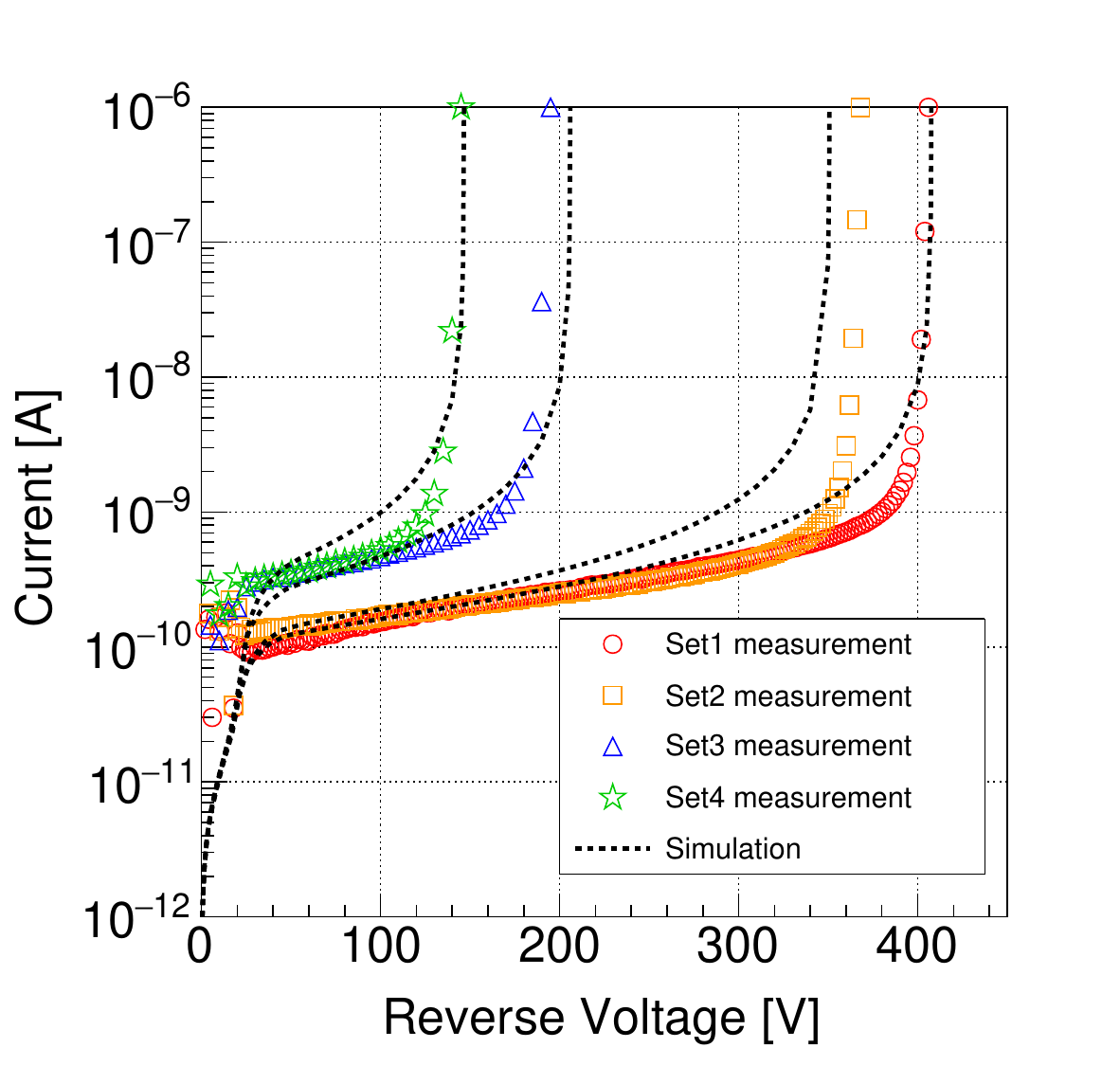}
        \caption{Leakage currrent for simulation (dashed line) and measurement for 4 different process sets after calibration. The details of 4 process sets are listed in \tablename~\ref{tab:process_set}.}
        \label{fig:4sets_iv}
    \end{figure}

We compare four different sets of simulated and measured I-V characteristics of LGAD sensors after calibration (Fig.\ref{fig:4sets_iv}) to verify the reliability of the process simulation setup, which is applied to all sets. The process sets are listed in \tablename~\ref{tab:process_set}. The LGAD sensor considered in Section~\ref{sec:ihep_ime_v1} belongs to set3 with a high implantation energy of phosphorus and high implantation dose of boron. All simulated breakdown voltages show good consistency with measurements. It proves that the calibrated configurations in the process simulation are reliable and $ V_{BD}$ could be reproduced by using the appropriate physical models.

    \begin{table*}[]

        \caption{LGAD process sets. Two implantation energy of phosphorus, Low / High that corresponding ratio are 1.00 / 1.25 and two implantation dose of boron, Low / High that corresponding ratio are 1.00 / 1.14.} 
        \centering
        \begin{tabular}{p{140pt}p{80pt}p{80pt}}
            \hline
            \hline
            \diagbox{Boron}{Phosphorus}       & Low Energy (1.00)   & High Energy (1.25)\\
            \hline
            Low Dose (1.00)     &  set2    &    set1        \\
            \hline
            High Dose (1.14)    &  set4     &   set3        \\  
            \hline       
        \end{tabular}
        \label{tab:process_set}
    \end{table*}

\section{Radiation model} \label{sec:rad_model}

\subsection{LGAD Radiation Damage Model (LRDM)}
Our LRDM TCAD model combines local acceptor removal with global deep energy levels (the Hamburg Penta Trap Model - HPTM \cite{Schwandt_HPTM}) in TCAD simulation to predict the current and capacitance of irradiated LGAD sensors. The acceptor removal takes effect in the gain layer whereas the deep energy levels are implemented in the high resistivity bulk region.  

The Poisson equation, which describes the acceptor removal and deep energy levels in the different regions is given by     

    \begin{small}
    \begin{equation}
        \label{eq:posisson}
         -\nabla \cdot (\varepsilon \nabla \phi) =q(p-n- \{ N_{A}^{eff} \}_{gain})+ \{ q\sum_{i} N_{t i}(\delta_{i}-f_{t i})\}_{bulk}
    \end{equation}
    \end{small}
    
    \noindent where $ N_{t i}$ is the trap concentration, $ \delta_{i}=1 (0)$ for donor (acceptor) type of trap and $ f_{t i}$ is the trap occupancy for trap i.

    In the acceptor removal model, the effective doping $ {N_{A}^{eff}}$ after irradiation is given by \cite{Ar_moll}:

    \begin{small}
    \begin{equation} \label{eq:ar}
         {N_{A}^{eff}}(\Phi_{eq}) = N_{A}^{eff}(0) + g \Phi_{eq} - N_{A}(0)[1- exp(-c \Phi_{eq})]
    \end{equation}
    \end{small}

\noindent where $ \Phi_{eq}$ is equivalent to 1 MeV neutron fluence by NIEL hypothesis, and $ N_{A}(0)$ and $ N_{A}^{eff}(0)$ are the initial and the effective doping without irradiation, respectively. We set the removal constant to $ c= 3.23 \times 10^{-16} ~cm^{2}$ by best-fitting and $ g = 0.02 ~cm^{-1}$ \cite{Ar_Model}. 

Concerning the deep energy levels at low temperature, we assume the validity of the HPTM, since it gives a consistent description of a large set of measurements in PIN with protons fluence range from $ 3 \times 10^{14} $ to $ 1.3 \times 10^{16} ~n_{eq}/cm^{2}$. The 5 deep energy levels of HPTM are implemented in silicon material with uniform distribution in TCAD simulations. 
   
In order to achieve the best fitting results, the introduction rate $ g_{int}(I_{p})$ of trap level $ I_{p}$ ($ E_{C}-0.545~eV$) for HPTM is set to 0.6050~$cm^{-1}$ (default is 0.4335~$cm^{-1}$) in our simulations.

\subsection{Simulated I-V and C-V characteristics of irradiated LGAD sensors}

Once we have confirmed our simulation setup and implemented the LRDM, Eq.(\ref{eq:posisson}) and the current continuity equations may be solved and the leakage current and capacitance of irradiated LGAD sensors may be estimated. The studied LGAD sensor has been irradiated at the Jozef Stefan Institute (JSI) with a flux of $ 2.5 \times 10^{15} ~n_{eq}/cm^{2}$. The measurements are performed at -30~$^{\circ}C$ after 60~$^{\circ}C$ for 80 minutes annealing.
    
\begin{figure*}[htb]
        \centering
        \subfigure[]{ \label{fig:rad_iv}
        \includegraphics[scale=0.33]{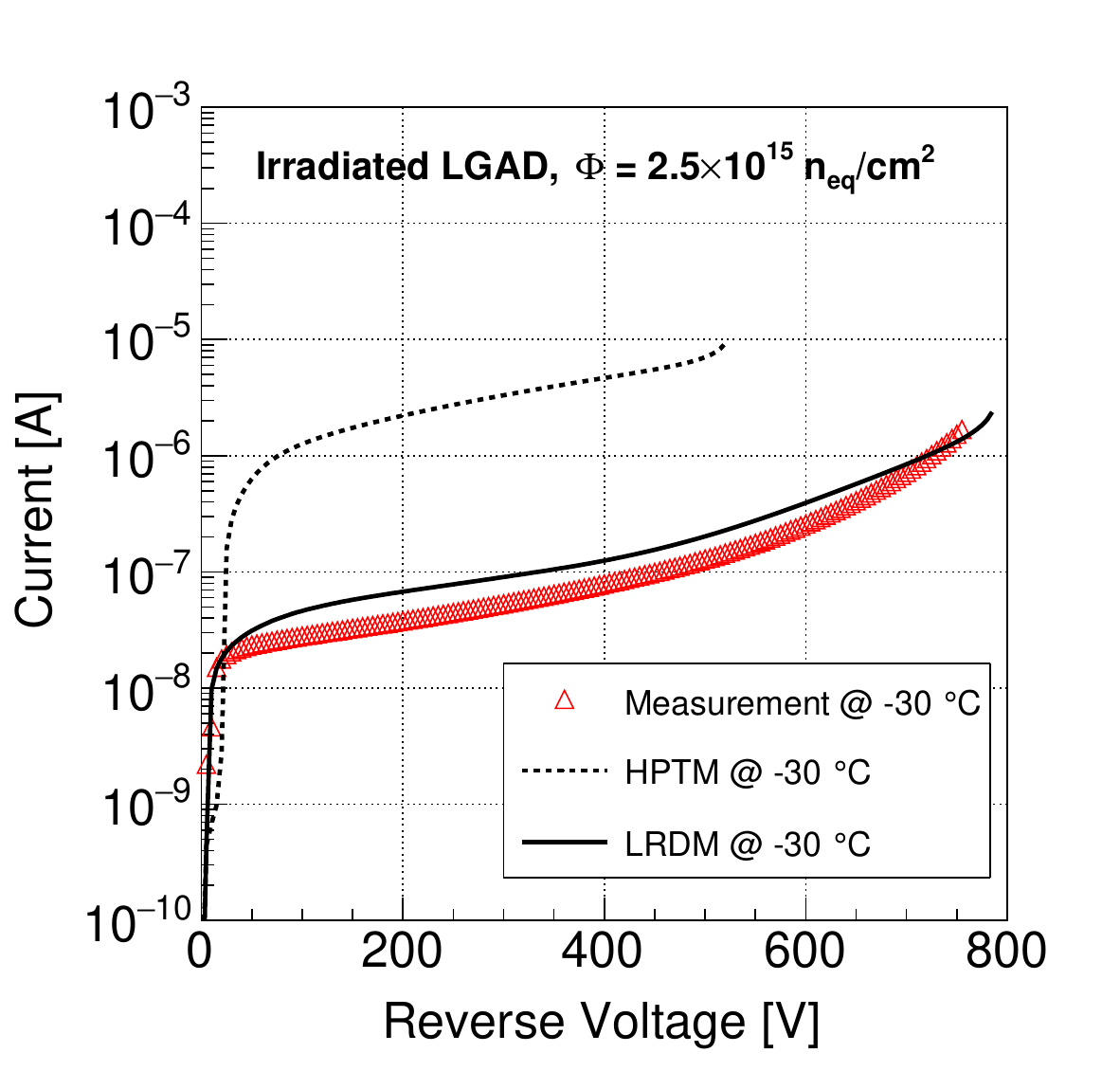}} 
        \subfigure[]{ \label{fig:rad_cv}
        \includegraphics[scale=0.33]{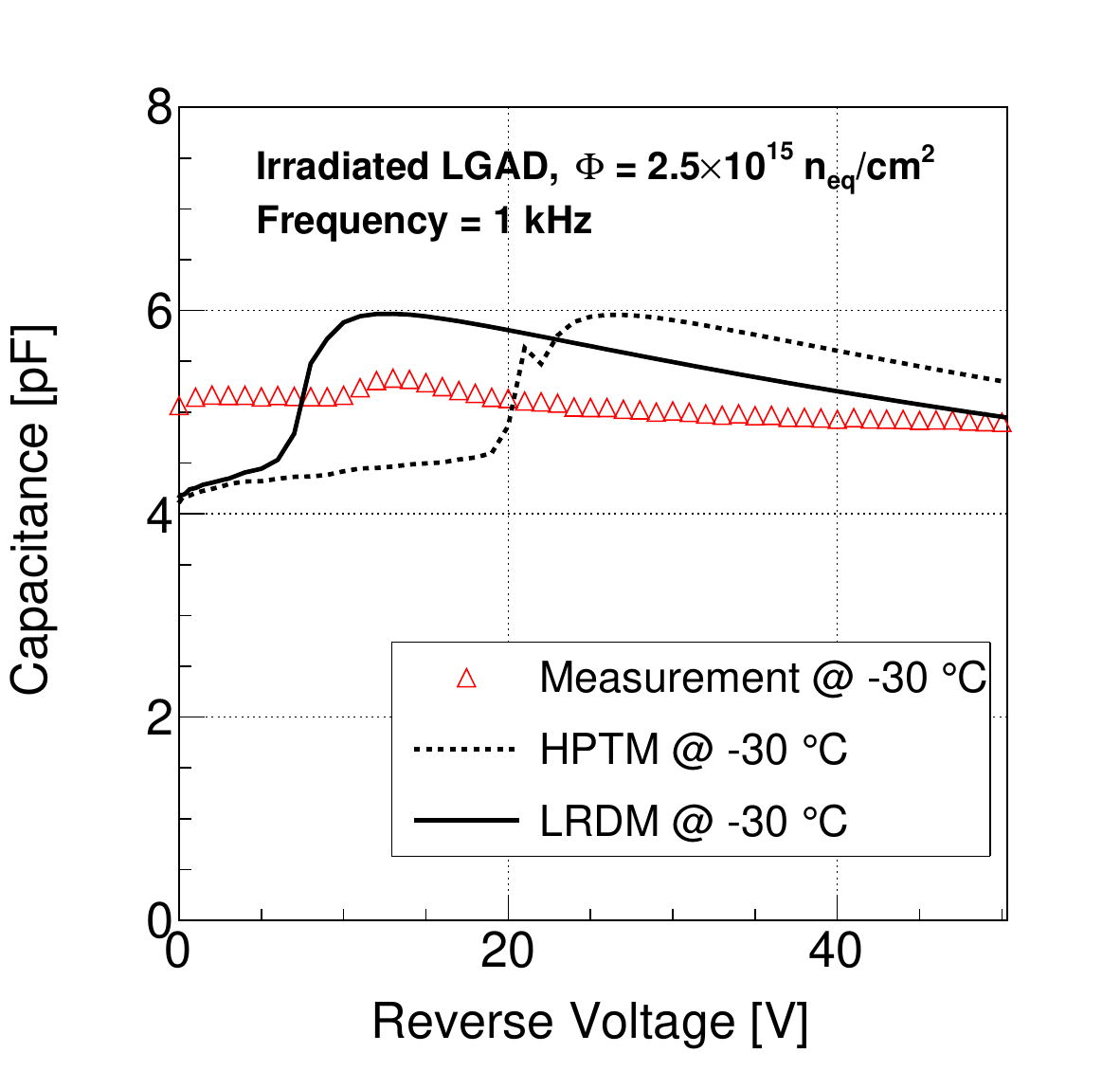}}
        \caption{Simulated (solid \& dashed line) and measured (triangle) (a) leakage currrent (b) capacitance for $ \Phi_{eq} = 2.5 \times 10^{15} n_{eq}/cm^{2}$ at -30 $ ^{\circ}C$.
        }
    \end{figure*}

The leakage current is determined by the residual gain and the deep energy traps as follows \cite{LGAD_Rad_Current}:

    \begin{small}
        \begin{equation} \label{eq:rad_current}
             I(\Phi_{eq}) = M_{I}(\Phi_{eq}) \times I_{gen}(\Phi_{eq})
        \end{equation}
    \end{small}

\noindent where $ M_{I}(\Phi_{eq})$ is dominated by the acceptor removal and $ I_{gen}(\Phi_{eq})$ by the deep energy levels. The simulated currents by HPTM and LRDM for $ \Phi_{eq} = 2.5 \times 10^{15} ~n_{eq}/cm^{2}$ are shown in Fig.\ref{fig:rad_iv}. Simulation by LRDM predicts a large increase of the leakage current and $ V_{BD}$ in irradiated LGAD sensors, which is consistent with measurements. The discrepancies between simulations and measurements are due to the uncertainty in the irradiation fluence and to temperature fluctuations. 

Fig.\ref{fig:rad_cv} illustrates the LRDM could reproduce the capacitance of irradiated LGAD to a certain degree compared with HPTM. The 1~kHz frequency is applied both in simulation and measurement after irradiation. When the fluence up to $2.5 \times 10^{15} n_{eq}/cm^{2}$, the gain layer is almost disappeared. Under this condition, the device is like a PIN, so the capacitance drops from  $\sim$400~pF to $\sim$6~pF. The behavior of capacitance at low temperature has been reported in previous works for irradiated CNM LGAD sensors \cite{CNM_LT_CV}. Comparing the extensive measurements \cite{Ar_Model,HPK_CV} for irradiated LGAD sensors at room temperature and low temperature, the capacitance near $V_{GL}$ has a ``bump'' only at low temperature. The estimated ``bump'' position voltages are 12.5~V by LRDM, 26.0~V by HPTM and 13.0~V by measurement, where the simulated results by LRDM agree with measurement. However, the theoretical interpretation about the appearance of ``bump'' is not complete, and the possible cause of the behavior of capacitance at low temperature is related to the complex space charge distribution (such as multi-junction) in irradiated LGAD, which is under investigation.
 
\section{Summary}
In this paper, we have presented and verified a full process simulation, including calibration, for the production of IHEP-IME LGAD sensors. Channeling suppression 
effects have been considered in process simulation using SIMS data and results
are in agreement with experiments. We obtain highly consistent characteristics for 
non-irradiated LGAD sensors, which indicates that our TCAD simulation setup may be reliably employed in the optimization of IHEP-IME-v2 production. It also provides solid ground for simulations of irradiated LGAD sensors. 

We have put forward an LRDM TCAD model, which combines local acceptor removal and global deep energy levels, to simulate the characteristics of irradiated LGAD sensors. The simulated leakage current well agrees with experimental results. The LRDM could qualitatively describe the behavior of capacitance at low temperature but need further investigation about capacitance levels.

Our results also pave the way for further studies about the radiation model and the theoretical description of the capacitance in irradiated LGAD sensors. However, present LRDM is still under development and it could not simulate the charge collection efficiency. And those studies will need more measurements supported by other techniques, e.g. deep energy levels in irradiated LGAD may be extracted by Thermally Stimulated Current (TSC) \cite{LGAD_Rad_TSC} and the electric field distribution may be obtained by edge -Transit Current Technique (e-TCT) \cite{LGAD_Rad_eTCT_Annealing} or Two-Photon Absorption - Transient Current Technique (TPA-TCT) \cite{LGAD_Rad_eTCT}. 

\section*{Acknowledgment}
This work has been supported by the National Natural Science Foundation of China (No. 11961141014), the Scientific Instrument Developing Project of the Chinese Academy of Sciences - Grant (No. ZDKYYQ20200007), and the National Key Research \& Development Program (No. 2016YFA0201903). We acknowledge key suggestions from the ATLAS HGTD Collaboration and CERN RD50 Collaboration and thank Joern Schwandt for his comments about the radiation model in TCAD simulations.

\bibliographystyle{unsrt}
\bibliography{p11_lgad_rad_sim}

\end{document}